\documentclass[prb,aps,amssymb,showpacs,twocolumn]{revtex4}
\usepackage{amsmath}
\usepackage{amssymb}
\usepackage{amsthm}
\usepackage{amsfonts}
\usepackage{enumerate}
\usepackage{latexsym}
\usepackage{graphicx}

\newcommand{\beq}{\begin{equation}}
\newcommand{\eneq}{\end{equation}}

\input{epsf}

\begin{document}

\tolerance 10000

\newcommand{\vk}{{\bf k}}


\title{Dynamical Axion Field in Topological Magnetic Insulators}
\author{Rundong Li$^1$, Jing Wang$^{2,1}$, Xiaoliang Qi$^1$ \& Shou-Cheng Zhang$^1$}

\affiliation{$^1$ Department of Physics, McCullough Building,
Stanford University, Stanford, CA 94305-4045} \affiliation{$^2$
Department of Physics, Tsinghua University, Beijing 100084, China}

\begin{abstract}
Axions are very light, very weakly interacting particles postulated
more than 30 years ago in the context of the Standard Model of
particle physics. Their existence could explain the missing dark
matter of the universe. However, despite intensive searches, they
have yet to be detected. In this work, we show that magnetic
fluctuations of topological insulators couple to the electromagnetic
fields exactly like the axions, and propose several experiments to
detect this dynamical axion field. In particular, we show that the
axion coupling enables a nonlinear modulation of the electromagnetic
field, leading to attenuated total reflection. We propose a novel
optical modulators device based on this principle.
\end{abstract}

\date{\today}


\maketitle

The electromagnetic response of a three dimensional insulators is
described by the Maxwells action $ S_0=\frac1{8\pi}\int d^3xdt
(\epsilon {\bf E}^2 - \frac{1}{\mu} {\bf B}^2)$, with
material-dependent dielectric constant $\epsilon$ and magnetic
permeability $\mu$, where ${\bf E}$ and ${\bf B}$ are the
electromagnetic fields inside the insulator. However, generally, it
is possible to include another quadratic term in the effective
action $S_\theta=\frac{\theta}{2\pi}\frac{\alpha}{2\pi} \int d^3xdt
{\bf E \cdot B}$, where $\alpha=e^2/\hbar c$ is the fine structure
constant, and $\theta$ is a parameter describing the insulator in
question. In the field theory literature this effective action is
known as the axion electrodynamics~\cite{wilczek1987}, where
$\theta$ plays the role of the axion field. Under the periodic
boundary condition the partition function and all physical
quantities are invariant if $\theta$ is shifted by integer multiples
of $2\pi$. Therefore all time reversal invariant insulators fall
into two distinct classes described by either $\theta=0$ or
$\theta=\pi$~\cite{Qi2008}. Topological insulators are defined by
$\theta=\pi$ and can only be connected continuously by time reversal
breaking perturbations to trivial insulators defined by $\theta=0$.
The form of the effective action implies that an electric field can
induce a magnetic polarization, whereas a magnetic field can induce
an electric polarization. This effect is known as the topological
magneto-electric effect (TME) and $\theta$ has the meaning of the
magneto-electric polarization $P_3=\theta/2\pi$. Physically the
parameter $\theta$ depends on the band structure of the insulator
and has a microscopic expression of the momentum space Chern-Simons
form~\cite{Qi2008}
\begin{equation}
\theta=\frac{1}{4\pi}\int{d^3k}\epsilon^{ijk}Tr\left[A_i\partial_jA_k+\frac{2}{3}A_iA_jA_k\right],\label{Theta}
\end{equation}
where
$A^{\alpha\beta}_i({\bf{k}})=-i\langle\alpha{\bf{k}}|\frac{\partial}{\partial{k_i}}|\beta{\bf{k}}\rangle$
is the momentum space non-abelian gauge field, with indices
$\alpha,\beta$ referring to the occupied bands. The $\theta$
parameter has been calculated explicitly for several basic models of
topological insulators\cite{Qi2008,Essin2009}. In a topological
insulator the axion field gives rise to novel physical effects such
as the image monopole and anyonic statistics \cite{qi2009}. This
field, however, is static in a time-reversal invariant topological
insulator. In this work, we consider the anti-ferromagnetic long
range order in a topological insulator, which breaks time-reversal
symmetry spontaneously, so that $\theta$ becomes a dynamical axion
field taking continuous values from $0$ to $2\pi$. In the following
we will refer to such an antiferromagnetic insulator as a
``topological magnetic insulator". We propose a minimal model in
which the antiferromagnetic order break the time reversal symmetry
spontaneously and the magnetic fluctuations couple linearly to the
axion field, thus realizing the dynamic axion field in condensed
matter systems. Compared to its high energy version, the axion
proposed here has the advantage that it can be observed in
controlled experimental settings \cite{Wilczek2009}. With an
externally applied magnetic field, the axion field couples linearly
to light, resulting in the axionic polariton. By measuring the
attenuated total reflection, the gap in the axionic polariton
dispersion can be observed. An attractive feature is that the
axionic polariton gap is tunable by changing the external electric
or magnetic fields. The control of the light transmission through
the material enables a novel type of optical modulator. We also
propose another experiment to detect the dynamic axion by
microcantilever torque magnetometry, where the double frequency
response of the cantilever is a unique signature of the dynamic
axion field.

We propose several materials that may realize the topological
magnetic insulator with dynamic axion field. One possibility is the
topological insulator $\rm{Bi_2Te_3}$, $\rm{Bi_2Se_3}$,
$\rm{Sb_2Te_3}$ doped with $3d$ transition metal elements such as
$\rm{Fe}$~\cite{Zhou2006,Larson2008}. Another possible class of
material is the 5d transition metal compound $\rm{A_xBO_y}$ with $B$
and $A$ standing for some 5d transition metal and some alkali metal,
respectively. Electrons in 5$d$-orbital can have both strong spin
orbital coupling and strong interaction, which is ideal for the
realization of the topological magnetic insulator\cite{Nagaosa2009}.
We show that such a compound with the corundum structure may have a
topological magnetic insulator phase if the states closed to fermi
level are formed by $t_{2g}$ orbitals with total angular momentum
$J_{\mathrm{eff}}=1/2$ \cite{Kim2009}. The detail of this proposal
is beyond the goal of the present work, and will be presented in a
separate paper.\cite{wang2009} We also noticed two very recent works
on 5d transition metal compounds with pyrochlore structure, which
may also realize the topological magnetic insulator
phase.\cite{pesin2009,guo2009}

\paragraph*{Effective model for the 3D topological insulator}
Although all the physical effects discussed in this paper are
generic for any system supporting axionic excitation and do not rely
on a specific model, we would like to start from a simple model for
concreteness. We adopt the effective model proposed by Zhang {\it et
al} in Ref. \cite{Zhanghaijun2009} to describe topological
insulators $\rm{Bi_2Te_3}$, $\rm{Bi_2Se_3}$ and $\rm{Sb_2Te_3}$. The
low energy bands of these materials consist of a bonding and an
anti-bonding state of $p_z$ orbitals, labeled by
$|P2^{-}_z,\uparrow(\downarrow)\rangle$ and
$|P1^{+}_z,\uparrow(\downarrow)\rangle$, respectively. The generic
form of the effective Hamiltonian describing these four bands is
obtained up to quadratic order of momentum ${\bf k}$ in Ref.
\cite{Zhanghaijun2009}. Since a lattice regularization is necessary
for computing axion field $\theta$, in the present paper we start
from a lattice version of this model, with the Hamiltonian
\begin{eqnarray}
    H_0({\bf k})&=& \epsilon_0({\bf k}){\rm I}_{4\times 4}+\sum_{a=1}^5d_a({\bf
    k})\Gamma^a\label{eq:Heff}\\
d_{1,2,...,5}({\bf{k}})&=&\left(A_2\sin k_x,A_2\sin k_y,A_1\sin
k_z,\mathcal {M}({\bf{k}}),0\right)\nonumber
\end{eqnarray}
where $\epsilon_0({\bf{k}})=C+2D_1+4D_2-2D_1\cos k_z-2D_2\left(\cos
k_x+\cos k_y\right)$, $\mathcal {M}({\bf{k}})=M-2B_1-4B_2+2B_1\cos
k_z+2B_2\left(\cos k_x+\cos k_y\right)$, and the Dirac $\Gamma$
matrices have the representation
$\Gamma^{(1,2,3,4,5)}=\left(\sigma_x\otimes{s}_x,\sigma_x\otimes{s}_y,\sigma_y\otimes{I}_{2\times2},
\sigma_z\otimes{I}_{2\times2},\sigma_x\otimes{s}_z\right)$ in the
basis of
$(|P1^{+}_z,\uparrow\rangle,|P1^{+}_z,\downarrow\rangle,|P2^{-}_z,\uparrow\rangle,|P2^{-}_z,\downarrow\rangle)$.

\paragraph*{$\mathcal{P,T}$ breaking terms.}
The above Hamiltonian preserves both time reversal symmetry
$\mathcal{T}$ and parity $\mathcal{P}$.

\begin{figure}[h]
\includegraphics[width=2.2in,clip=true]{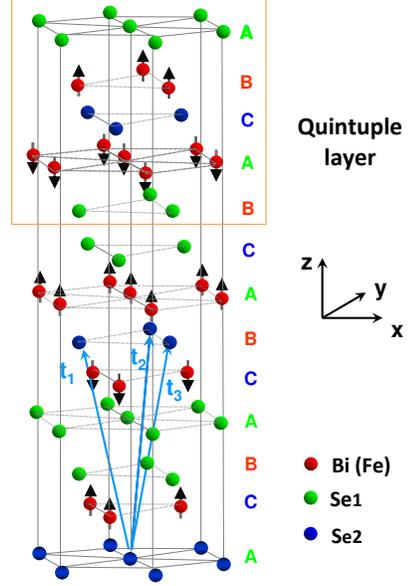}
\caption{ [Color Online] Crystal structure of $Bi(Fe)_{2}Se_3$ with
three primitive lattice vectors denoted as $\mathbf{t}_{1,2,3}$. A
quintuple layer with $Se1-Bi(Fe)1-Se2-Bi(Fe)1^{\prime}-Se1^{\prime}$
is indicated in the orange box. The spin ordering configuration
giving rises to the $\Gamma_5$ mass is indicated by the black arrow,
which is antiferromagnetic along $z$ direction and ferromagnetic
within $xy$ plane.} \label{structure}
\end{figure}

\noindent In the effective model (\ref{eq:Heff}), the time-reversal
and spatial inversion transformation are defined as
$\mathcal{T}=i(I_{2\times2}\otimes s_y)\cdot\mathcal{K}$ (with
$\mathcal{K}$ the complex conjugation operator) and
$\mathcal{P}=\sigma_z\otimes I_{2\times2}$, respectively. The axion
field $\theta$ can be calculated by formula (\ref{Theta}) which
gives $0$ or $\pi$ depending on the value of
parameters\cite{Qi2008}, as expected from time reversal symmetry.
Now we consider a perturbation to the Hamiltonian (\ref{eq:Heff})
which can lead to deviation of $\theta$ from $0$ or $\pi$. Since
$\theta$ is odd under time reversal and parity operation, only time
reversal and parity breaking perturbations can induce a change of
$\theta$. By simple algebra one can show that to the leading order
the $\mathcal{T,P}$ breaking perturbation must have the form $\delta
H({\bf k})=\sum_{i=1,2,3,5}m_i\Gamma^i$. Thus the perturbed
Hamiltonian can still be written as
$H({\bf{k}})=\sum\limits_{a=1}^5d_a({\bf{k}})\Gamma^a$  with
$d_a({\bf{k}})=\left(A_2\sin{k_x}+m_1,A_2\sin{k_y}+m_2,A_1\sin{k_z}+m_3,
\mathcal{M}({\bf{k}}),m_5\right)$. For this model, Eq.~(\ref{Theta})
can be reduced to an explicit expression for $\theta$:
\begin{equation}
\theta =\frac{1}{4\pi}\int{d}^3k
\frac{\left(|d|+d_4\right)\left(4|d|-d_4\right)}{\left(|d|^2+d_4|d|\right)^3}
d_5 (\partial_{k_x}d_1) (\partial_{k_y}d_2)
(\partial_{k_z}d_3).\label{Thetavsd}
\end{equation}
Although there are four independent parameters $m_{1,2,3,5}$ in
$\mathcal{T,P}$ breaking term, only $m_5$ leads to a correction to
$\theta$ to the linear order. Thus in the following we will take
$m_{1,2,3}=0$ without the loss of generality, and leave only the
$m_5\Gamma^5$ term. To see the physical meaning of the $m_5\Gamma^5$
term we change the basis to
$\left(|A,\uparrow\rangle,|A,\downarrow\rangle,|B,\uparrow\rangle,|B,\downarrow\rangle\right)$
with
$|A,\uparrow(\downarrow)\rangle=\frac{1}{\sqrt{2}}\left(|P1^{+}_{z},\uparrow(\downarrow)\rangle+|P2^{-}_z,\uparrow(\downarrow)\rangle\right)$
and
$|B,\uparrow(\downarrow)\rangle=\frac{1}{\sqrt{2}}\left(|P1^{+}_{z},\uparrow(\downarrow)\rangle-|P2^{-}_z,\uparrow(\downarrow)\rangle\right)$.
We see that
$\mathcal{P}|A,\uparrow(\downarrow)\rangle=|B,\uparrow(\downarrow)\rangle$.
Physically, $|A,\uparrow(\downarrow)\rangle$ and
$|B,\uparrow(\downarrow)\rangle$ stand for states of the two lattice
sites in each unit cell, {\it e.g.} the two ${\rm Bi}$ atom sites in
${\rm Bi_2Te_3}$, which are shifted away from the inversion center
along $z$ direction. By transforming $\Gamma^5$ to the new basis, we
see that it represents a staggered Zeeman field that points in
$+z(-z)$ direction on the $A(B)$ sublattice. This staggered Zeeman
field can be generated by an antiferromagnetic order, where electron
spins point along opposite $z$ directions on two sublattices, as
shown in Fig. 1. Without the $\Gamma^5$ term, topological insulators
have protected surface states consisting of odd number of massless
Dirac cones. With broken time reversal symmetry, a direct
consequence of the $m_5\Gamma^5$ term is that it opens a gap in the
surface state spectrum. The surface state gap is equal to $m_5$,
independent on the orientation of the surface.

\paragraph*{Dynamical axion in the anti-ferromagnetic phase} We have
seen above that the axion field $\theta$ can deviate from $0$ or
$\pi$ if the electrons are coupled to the antiferromagnetic order
parameter. Such a staggered field can be induced if an
antiferromagnetic long range order is established in the system.
Physically, an antiferromagnetic ordered phase can be obtained
naturally if the screened Coulomb interaction of electrons become
strong. For example, in the ${\rm Bi_2Te_3}$ family doped with
magnetic impurities such as ${\rm Fe}$, the substitution of
$p$-electrons of ${\rm Bi}$ or ${\rm Te}$  by $d$-electrons of the
magnetic impurities effectively enhanced the on-site repulsion of
electrons. In the basis
$\left(|A,\uparrow\rangle,|A,\downarrow\rangle,|B,\uparrow\rangle,|B,\downarrow\rangle\right)$
the Hamiltonian with interaction can be written as
\begin{equation}
H=H_0+U\sum_{i}\left(n_{iA\uparrow}n_{iA\downarrow}+n_{iB\uparrow}n_{iB\downarrow}\right)+V\sum_{i}n_{iA}n_{iB}.
\end{equation}
in which the first term is the kinetic energy given by Eq.
(\ref{eq:Heff}) and the rest two terms represent the on-site
repulsion $U$ and the inter-site repulsion $V$ between $A$ and $B$
sites. Possible ordered phase resulting from the interaction include
the ferromagnetic phase where the spin on two sublattices $A$ and
$B$ point in the same direction, the antiferromagnetic phase where
the spin on two sublattices point in the opposite direction, and the
charge density wave (CDW). Correspondingly the order parameters are
taken as the ferromagnetic order parameter
${\bf{M}}^{+}=\frac{1}{2}\left(\langle{\bf{S}}_{iA}\rangle+\langle{\bf{S}}_{iB}\rangle\right)$,
the antiferromagnetic order parameter
${\bf{M}}^{-}=\frac{1}{2}\left(\langle{\bf{S}}_{iA}\rangle-\langle{\bf{S}}_{iB}\rangle\right)$
and the CDW order parameter
$\rho=\frac{1}{2}\left(\langle{n}_{iA}\rangle-\langle{n}_{iB}\rangle\right)$.
It is assumed that translational symmetry is preserved and all the
order parameters are uniform in space. In the mean field
approximation, we find that for a wide range of value for band
structure parameters $M$, $A_{1,2}$ and $B_{1,2}$, the system
develops antiferromagnetic order pointing in the $z$ direction ${\bf
M^-}=M^-_0{\bf \hat{z}}$ if the effect of $U$ dominates that of $V$,
which thus leads to $m_5=-\frac23UM_z^-$ and axion field $\theta\neq
0,\pi$.

\paragraph*{Axion electrodynamics.} In the mean-field approximation,
the antiferromagnetic phase has a static axion field $\theta$.
However, the antiferromagnetic phase also has amplitude and spin
wave excitations, which can induce fluctuations of the axion field.
The fluctuation of the Neel vector ${\bf M}^-$ can be generally
written as ${\bf M}^-=\left(M_0^-+\delta M_z({\bf x},t)\right){\bf
\hat{z}}+\delta M_x({\bf x},t){\bf \hat{x}}+\delta M_y({\bf
x},t){\bf \hat{y}}$. To the linear order, it can be shown from
symmetry analysis that the fluctuation of axion field only depends
on $\delta M_z$, since $\theta$ is a pseudo-scalar. In other words,
we have $\delta\theta({\bf x},t)=\delta m_5({\bf x},t)/g=-\frac
23U\delta M_z({\bf x},t)/g$ where the coefficient $g$ can be
determined from Eq. (\ref{Thetavsd}). The dispersion of the
amplitude mode $\delta M_z({\bf x},t)$ can be obtained in standard
RPA approximation, leading to a massive axion field
$\delta\theta({\bf x},t)$. Considering the coupling term $\theta
{\bf E\cdot B}$ of axion with electromagnetic field, the effective
action describing the axion-photon coupled system is given by
\begin{eqnarray}
\mathcal{S}_{\mathrm{tot}} &=& \mathcal{S}_{\mathrm{Maxwell}} +
\mathcal{S}_{\mathrm{topo}} +\mathcal{S}_{\mathrm{axion}}\nonumber\\
&=&\frac1{8\pi}\int d^3xdt (\epsilon {\bf E}^2 - \frac{1}{\mu} {\bf
B}^2)\nonumber\\
&+&\frac{\alpha}{4\pi^2}\int
d^3xdt\left(\theta_0+\delta\theta\right)
{\bf{E}}\cdot{\bf{B}}\nonumber\\
&+&g^{2}J\int
d^3xdt\left[(\partial_{t}\delta\theta)^2-(v_i\partial_i\delta\theta)^2-m^2\delta\theta^2\right],\label{Action}
\end{eqnarray}
where $J,v_i$ and $m$ are the stiffness, velocity and mass of the
spin wave mode $\delta M_z$, $\bf{E}$ and $\bf{B}$ are the electric
field and the magnetic field respectively, $\epsilon$ and $\mu$ are
the dielectric constant and magnetic permeability respectively. The
second term describes the topological coupling between the axion and
the electromagnetic field, with $\alpha\equiv e^2/\hbar c$ the
fine-structure constant. The third term describes the dynamics of
the massive axion. Within the model we have adopted, the parameters
$J$ and $m$ are given by
\begin{eqnarray}
J&=&\int\frac{d^3k}{(2\pi)^3}\frac{d_i({\bf{k}})d^i({\bf{k}})}{16|d|^5}\nonumber\\
Jm^2&=&\left(\frac{2}{3}UM^-_z\right)^2\int\frac{d^3k}{(2\pi)^3}\frac{1}{4|d|^3},
\end{eqnarray}
where $|d|=\sqrt{\sum_{a=1}^5d_ad^a}$ and the repeated index
indicates summation with $i=1,2,3,4$.

\paragraph*{The axionic polariton.} The dynamic axion field $\theta$
couples nonlinearly to the external electromagnetic field
combination ${\bf{E}}\cdot{\bf{B}}$. When there is an externally
applied static and uniform magnetic field ${\bf{B}}_0$ parallel to
the electric field ${\bf{E}}$ of the photon, $\theta$ will couple
linearly to $\bf{E}$~\cite{Maiani1986}. In condensed matter systems,
when a collective mode is coupled linearly to photons, hybridized
propagating modes called polariton emerge\cite{mills1974}. The
polaritons can be coupled modes of optical phonon and light through
the electric dipole interaction, or coupled modes of magnon and
light through the magnetic dipole interaction. Here we propose a
novel type of polariton---{\em axionic polariton} which is the
coupled mode of light and the axionic mode of an antiferromagnet.
The dispersion of the axionic polariton can be obtained from the
effective action (\ref{Action}) which leads to the following
linearized equation of motion~\cite{Raffelt1988,Cameron1993}
\begin{eqnarray}\label{motion}
\frac{\partial^2}{\partial{t}^2}E-c'^2{\nabla}^2E+\frac{\alpha{B_0}}{\pi\epsilon}\frac{\partial^2}{\partial{t}^2}\delta\theta&=&0\nonumber\\
\frac{\partial^2}{\partial{t}^2}\delta\theta-v^2{\nabla}^2\delta\theta+m_0^2\delta\theta-\frac{\alpha{B_0}}{8\pi^2{g^2}J}E&=&0,
\end{eqnarray}

\begin{figure}[h]
\includegraphics[scale=0.40]{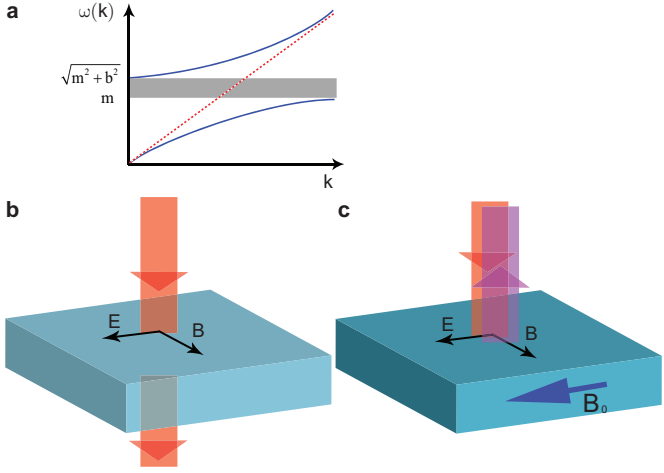}
\caption{ [Color Online] Axionic polariton and ATR experiment. {\bf
a.} The dispersion of the axionic polariton. The grey area indicates
the forbidden band between frequencies $m$ and $\sqrt{m^2+b^2}$ (see
text), within which light can not propagate in the sample. The red
dashed line shows the bare photon dispersion $\omega=c'k$. {\bf b.}
Setup for the attenuated total reflection (ATR) experiment. Without
external magnetic field, the incident light can transmit through the
sample. {\bf c.} When an external magnetic field is applied parallel
to the electric field of light, the incident light will be totally
reflected if its frequency lies within the forbidden band.}
\label{ATR}
\end{figure}

\noindent where $c'$ is the speed of light in the media and
$\epsilon$ is the dielectric constant. Compared to photon, the
dispersion of axion can be neglected, in which case the axionic
polaritons have the dispersion
\begin{eqnarray}
\omega^2_{\pm}(k)&=&\frac{1}{2}\left(c'^2k^2+m^2+b^2\right)\nonumber\\
&\pm&\frac{1}{2}\sqrt{\left(c'^2k^2+m^2+b^2\right)^2-4c'^2k^2m^2},\label{dispersion}
\end{eqnarray}
with $b^2=\alpha^2{B_0}^2/8\pi^3\epsilon{g^2}{J}$. As shown in Fig.
2 a, this dispersion spectrum consists of two branches separated by
a gap between $m$ and $\sqrt{m^2+b^2}$. The quantity $b$ measures
the coupling strength between the axion field and the electric field
and is proportional to the external magnetic field $B_0$. Upon
turing on $B_0$, the axionic mode at $k=0$ changes its frequency
from $m$ to $\sqrt{m^2+b^2}$, due to the linear mixing between the
axion and the photon field. Physically, the axionic polariton is
very similar to the transverse optical phonon polariton, since the
axion also leads to an additional contribution to the charge
polarization due to the topological magneto-electric
effect\cite{Qi2008}, ${\bf P}=\alpha {\bf B}_0/\pi\theta+\epsilon
{\bf E}$. The optical phonon polariton has the same dispersion as
Eq. (\ref{dispersion}), with the parameter $b$ replaced by the
lattice unscreened plasmon frequency
$\omega_p=\sqrt{\frac{4\pi{n}e^{*2}}{m^*}}$. The key difference
between axion and optical phonon is that the coupling between axion
and electric field is determined by the external magnetic field
${\bf B}_0$, which is thus {\em tunable}.

The gap in the axionic polariton spectrum may be experimentally
observed using the attenuated total reflection (ATR) method. The
geometry is arranged such that the incident light is perpendicular
to the surface of the sample and the static magnetic field is
parallel to the electric field of light, as shown in Fig. 2 b and c.
Since the light can only propagate through the media in the form of
the axionic polariton, when the frequency of the incident light is
within the gap of the axionic polariton spectrum, a significant
increase of the reflectivity will be observed. To estimate the gap
we adopt the tight-binding parameters obtained for
$\rm{Bi_2Se_3}$\cite{Zhanghaijun2009} in Hamiltonian
(\ref{eq:Heff}). We take a typical exchange splitting for an
antiferromagnet $m_5=15{\rm meV}$, and an estimated dielectric
constant $\epsilon=100$. With a magnetic field $B_0=2T$, we obtain
the axion mass $m=36.7meV$ and $b=2.2meV$. The gap is approximately
$\frac{b^2}{2m}=0.25meV$, which can be observed experimentally. One
unique signature of the axionic polariton is the dependence of the
gap on $B_0$, which can be used to distinguish from usual magnetic
polaritons.

\begin{figure}[h]
\includegraphics[scale=0.36]{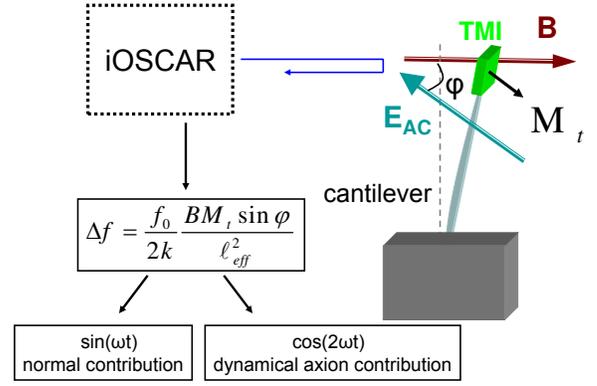}
\caption{ [Color Online] Cantilever Torque Magnetometry measurement
of axion. A cantilever experiences a torque $\vec{\tau}$ when the
magnetization ${\bf M}_{t}$ of the deposited topological magnetic
insulator induced by the AC electric field ${\bf E}(t)$ is placed in
the applied magnetic field ${\bf B}$. The $\omega$ and $2\omega$
frequency response comes from normal contribution and dynamical
axion contribution, respectively. ``iOSCAR'' is so called
interrupted oscillating cantilever-driven adiabatic reversal
protocol which can be used to measure the frequency change of
cantilever\cite{Budakian2003}.} \label{cantilever}
\end{figure}

\noindent By changing the magnitude of $B_0$ we can selectively
determine the frequency band within which the light is totally
reflected. This principle may find application as an amplitude
optical modulator working at far-infrared frequency.

\paragraph*{Measuring the axion by microcantilever.}

As discussed in Ref. \cite{Qi2008}, a static $\theta$ leads to a
topological magneto-electric effect, in which a magnetization
$\mathbf{M}_t=-\frac{\alpha}{4\pi^2}\theta\mathbf{E}$ is induced by
electric field. When the dynamics of axion field $\theta$ is
considered, the behavior of the magnetization ${\bf M}_t$ is
modified, which can be detected by microcantilever torque
magnetometry (MTM).~\cite{Lohndorf2000,Budakian2003,Budakian2005} As
shown in Fig. 3, a DC magnetic field $\mathbf{B}$ and an AC electric
field ${\bf E}(t)={\bf E}_{AC}\sin(\omega t)$ are applied to the
topological insulator attached to the tip of the cantilever, with an
angle $\varphi$ between them. A magnetization $\mathbf{M}_t$ will be
induced along the direction of the electric field, so that a torque
$\vec{\tau}=\mathbf{M}_t\times\mathbf{B}$ acting on the cantilever
is generated. This magnetic force on the cantilever mimics a change
in cantilever stiffness, in turn, shifts the cantilever frequency by
$\Delta
f=\frac{f_0}{2k}\frac{M_tB\sin\varphi}{(l_{\rm{eff}})^2}$~\cite{Budakian2003,Budakian2005},
Here $k$ is the cantilever stiffness, $f_0$ is the cantilever
natural frequency, $l_{\rm{eff}}$ is the effective length of
cantilever. With the electric field $\mathbf{E}_{AC}\sin(\omega t)$,
the time dependent frequency change is,
\begin{equation}
\Delta f_{AC}(t) = a_1\sin(\omega t)+a_2\cos(2\omega t),
\end{equation}
with
\begin{eqnarray}
a_1 &=&
\frac{\alpha}{4\pi^2}\frac{f_0}{2k(l_{\rm{eff}})^2}B\theta_0E_{AC}\sin\varphi,\nonumber
\\
a_2 &=& \frac{\alpha^2}{32\pi^4}\frac{f_0}{2k(l_{\rm{eff}})^2}\frac{
B^2E_{AC}^2\sin2\varphi}{4g^2J(m^2-\omega^2)}.
\end{eqnarray}
It should be noticed that the response $a_2$ with doubled frequency
$2\omega$ is a unique signature of the axion dynamics, since all the
response should have been linear and thus in frequency of $\omega$
if the axion field $\theta$ is static. With typical cantilever
parameters in present experiments~\cite{Budakian2003,Budakian2005}
$f_0=5.3$kHZ, $k=2.0\times10^{-4}$N/m, $l_{\rm{eff}}=500\mu$m,
$B=10^3$G, $E_{\rm{AC}}=10^4$V/m, $\omega/2\pi=1.5$Hz and
$\varphi=\pi/4$, we obtain the frequency shift $a_1=120$Hz and
$a_2=4.04$Hz, which are easily detectable in current experimental
techniques.

In conclusion, we have proposed the existence of dynamical axions in
topological magnetic insulators. We have proposed two experiments in
which axions can be detected by its unique coupling to the
electromagnetic field. The coupling between axion and photon leads
to an axionic polariton which has a polariton gap tunable by
magnetic field and thus may be used as a tunable optical modulator.

We wish to thank T. L. Hughes, S. B. Chung, S. Raghu, J.
 Maciejko, R. B. Liu and B. F. Zhu for insightful discussions. This work is supported by
the US Department of Energy, Office of Basic Energy Sciences under
contract DE-AC03-76SF00515. JW acknowledges the support of China
Scholarship Council, NSF of China (Grant No.10774086), and the
Program of Basic Research Development of China (Grant No.
2006CB921500).

\end{document}